\documentclass[aps,prl,twocolumn,nofootinbib,showpacs]{revtex4}

\usepackage{epsf,epsfig,amsmath,amssymb,dsfont}

\newsavebox{\ns}
\newsavebox{\dbrane}

\def\be{\begin{equation}}
\def\ee{\end{equation}}
\def\ben{\begin{equation*}}
\def\een{\end{equation*}}
\def\bea{\begin{eqnarray}}
\def\eea{\end{eqnarray}}
\def\bal{\begin{align}}
\def\eal{\end{align}}

\def\Dslash{\,\,{\raise.15ex\hbox{/}\mkern-12mu D}}
\def\Dbarslash{\,\,{\raise.15ex\hbox{/}\mkern-12mu {\bar D}}}
\def\delslash{\,\,{\raise.15ex\hbox{/}\mkern-9mu \partial}}
\def\delbarslash{\,\,{\raise.15ex\hbox{/}\mkern-9mu {\bar\partial}}}
\def\pslash{\,\,{\raise.15ex\hbox{/}\mkern-9mu p}}
\def\calDslash{\,\,{\raise.15ex\hbox{/}\mkern-12mu {\cal D}}}

\newcommand\CS{\mathcal{C}}

\newcommand{\mbb}{\mathbb}

\newcommand{\nn}{\nonumber \\}

\newcommand{\fr}{\frac}

\begin{document}

\title{The Scalar Curvature of a Causal Set}

\author{Dionigi M. T. Benincasa and Fay Dowker}
\affiliation{Blackett Laboratory,
Imperial College, London SW7 2AZ, United Kingdom}

\begin{abstract}

A one parameter family of retarded linear operators on scalar fields 
on causal sets is introduced. When the causal set is 
well approximated by 4 dimensional Minkowski 
spacetime, the operators are Lorentz invariant 
but nonlocal, are  parametrised 
by the scale of the nonlocality and approximate the
continuum scalar D'Alembertian $\Box$ when acting on fields that 
vary slowly on the nonlocality scale. 
The same operators can be applied to scalar fields on 
 causal sets which are well approximated by curved spacetimes
in which case they approximate $\Box - {\frac{1}{2}}R$ 
where $R$ is the Ricci scalar curvature. 
This can used to define an approximately
local action functional for causal sets.

\end{abstract}

\maketitle

The coexistence of Lorentz symmetry and 
fundamental, Planck scale spacetime discreteness has its
price: one must give up locality.
Since, if our spacetime is granular at the Planck scale, 
the ``atoms of spacetime'' that are nearest 
neighbours to a given atom will be of order one Planck 
unit of proper time away from it. The locus of such points
in the approximating continuum Minkowski spacetime 
is a hyperboloid of infinite spatial 
volume on which Lorentz transformations 
act transitively. The nearest neighbours 
will, loosely, comprise this hyperboloid and so there will 
be an infinite number of them. Where curvature limits 
Lorentz symmetry, it may render the number of nearest neighbours 
finite but it will still be huge so long as the radius of curvature
is large compared to the Planck length. 
Causal set theory 
is a discrete approach to 
quantum gravity which embodies Lorentz 
symmetry  \cite{Bombelli:1987aa,Bombelli:2006nm} 
and exhibits nonlocality of
exactly this form \cite{Moore:1988zz, Bombelli:1988qh}.

Nonlocality  
looks to be simultaneously a blessing and
a curse in tackling the twin challenges
that any fundamentally discrete 
approach to the problem of quantum gravity must face.
These are to explain (1) 
how the fundamental dynamics picks out
a discrete structure that is well approximated by a Lorentzian manifold
and (2) why, in that case, the geometry should be a solution
of the Einstein equations. This is often referred to 
as the problem of the continuum limit but in the context of 
a fundamentally discrete theory in which the 
discreteness scale is fixed and is not taken to zero
but rather the {\emph{observation scale}} is large, 
it is more accurately described 
as the problem of the continuum approximation. 

Consider first the problem of recovering 
a continuum from a quantum theory of {\emph{discrete manifolds}}.
(We adopt this term following Riemann \cite{Riemann:1868} and use 
it to refer to causal sets, simplicial complexes, graphs, or
whatever discrete entities the underlying theory is based on.)
Whenever a background principle or structure in a physical theory 
is abandoned in order to seek a {\emph{dynamical}} explanation for 
that structure, the state we actually 
observe becomes a very special one amongst 
the myriad possibilities that 
then arise. The continuum is just such a background 
assumption. In giving it up, generally one 
introduces a space of discrete manifolds in which 
the vast majority have no
continuum approximation. There will therefore be a 
competition between 
the entropic pull of the huge number of noncontinuum 
configurations -- 
choose one 
uniformly at random and it will not look anything like our spacetime --
and the dynamical law which must suppress the contributions 
of these nonphysical configurations to the path integral. 
The following general argument shows that a {\emph{local}} dynamics 
for quantum gravity will struggle to provide the required suppression. 
Consider the partition function as 
a sum over histories in which the
weight of each discrete manifold is $e^{-S}$
where $S$ is the real Wick rotated action.
As we increase the observation scale, 
the sum will be over discrete manifolds with an
increasing number, $N$, of atoms.  
If the action is local -- which in a discrete setting
translates to it being a sum over
contributions from each atom -- 
then it will grow no faster than 
$N$ times some constant, $\alpha$, and so
each weight is no smaller than $e^{-\alpha N}$.
If the number of 
discrete manifolds with $N$ atoms grows faster than exponentially  
with $N$, and if the majority of these discrete manifolds are
not continuumlike then 
they will overwhelm the partition 
function and the typical configuration will not have a continuum 
approximation. Even when the number of discrete manifolds 
is believed to grow exponentially, 
entropy can still trump 
dynamics as was seen in the lack of a continuum 
limit in the Euclidean 
dynamical triangulations programme \cite{Agishtein:1991cv,
Agishtein:1992xx, Ambjorn:1991pq, Ambjorn:1994tza}. 
Causal dynamical triangulations
do better, see, {\emph{e.g.}}, 
\cite{Ambjorn:2004qm,Ambjorn:2005db,Ambjorn:2007jv,Ambjorn:2008wc}, 
by restricting the class of triangulations allowed in the
sum.

In the case of causal sets, the number of 
discrete manifolds of size $N$ grows
as $e^{N^2/4}$ 
\cite{Kleitman:1975} and a local action would give causal set 
theory little chance of recovering the continuum. 
So the nonlocality of causal sets
holds out hope that the theory has a continuum regime and
indeed there exist physically motivated, classically stochastic
dynamical models for causal sets \cite{Rideout:1999ub}
in which the entropically
favoured configurations almost surely do not occur and those
that do exhibit an intriguing hint of manifold-like-ness
\cite{Ahmed:2009qm}.

However, nonlocality poses a danger when it comes to 
the second challenge of recovering
Einstein's equations. 
If we assume that a discrete quantum gravity theory does have
a 4 dimensional continuum regime, and if the theory is 
local and generally covariant, then the 
long distance physics 
will be governed by an effective Lagrangian which is 
a derivative expansion in which all 
diffeomorphism invariant terms are present but
higher derivative terms
are suppressed by the appropriate powers of the 
Planckian discreteness length scale, $l$:
\be
\frac{{\cal{L}}_{\textrm{eff}}}{\sqrt{-g}\hbar} 
= a_0 l^{-4} + a_1 l^{-2} R + a_2 R^2 + \dots
\ee
where $R$ is the Ricci scalar,
$a_1$ and $a_2$ are dimensionless couplings of 
order 1, and the dots denote further curvature squared terms 
as well as cubic and higher terms. 
The coefficient of
the leading term, $a_0$, is also naturally of order 1 which would 
make it 120 orders of  magnitude larger than its observed value.
However, that would also produce curvature on Planckian 
scales and so would not be compatible with the assumption of
a continuum approximation. In a discrete theory,
the question of why the cosmological constant 
does not take its natural value 
is the same question as why there is a continuum regime at all
and we must look to the fundamental dynamics for its resolution.
Assuming there is a resolution and a continuum regime
exists, locality and general covariance then pretty much 
guarantee Einstein's equations due to the natural suppression  
of the curvature squared and higher terms compared to the 
Einstein-Hilbert term.
     
So, Lorentz symmetry and 
discreteness together imply nonlocality, but 
nonlocality blocks the recovery of
general relativity, 
and if causal sets were incorrigibly nonlocal, this 
would be fatal. Suppose, however, that the 
nonlocality were somehow limited to 
length scales shorter than a certain $l_k$, 
which could be much larger than the 
Planckian discreteness scale, $l$, but yet have remained experimentally 
undetected to date.  There is already evidence that this 
is possible and indeed causal sets
admit constructions that are 
local enough to approximate the scalar D'Alembertian 
operator in 2 dimensional flat spacetime \cite{Henson:2006kf,
Sorkin:2007qi}. We add to this evidence here by exhibiting 
a family of discrete operators that approximate the scalar 
D'Alembertian in 4 dimensional flat spacetime. 
Further, both the 2D and 4D operators, when 
applied to scalar fields on causal sets which 
are well described by curved spacetimes 
approximate $\Box - \frac{1}{2} R$, where $R$ is the 
Ricci scalar curvature. We use this to propose an
action for a causal set which is approximately local.

We recall that a \emph{causal set} (or \emph{causet}) is 
a locally finite partial order, \emph{i.e.},
 it is a pair $(\CS,\preceq)$ where $\CS$ is a set 
and $\preceq$ is a partial order relation on $\CS$, which is
(i) reflexive: $x\preceq x$, 
(ii) acyclic $x\preceq y\preceq x \Rightarrow x=y$, and 
(iii) transitive $x\preceq y\preceq z\Rightarrow x\preceq z$,
for all $x,y,z\in \CS$. Local finiteness is the condition that 
the cardinality of any {\emph{order interval} is finite, 
where the (inclusive) order interval between a pair of elements
$y\preceq x$ is defined to be
$I(x,y) := \{z\in \CS \,|\, y\preceq z \preceq x\}$.
We write $x \prec y$ when $x \preceq y$ and $x \ne y$.
We call a relation $x\prec y$ a \emph{link} 
if the order interval $I(x,y)$ contains only $x$ and $y$: 
they are nearest neighbours. 
 
\emph{Sprinkling} is a way of generating a causet from a 
$d$-dimensional Lorentzian manifold $(\mathcal{M},g)$. 
It is a Poisson process of selecting points in $\mathcal{M}$
with density $\rho$ so that the  expected number of points 
sprinkled in a region of spacetime volume $V$ is $\rho V$. 
This process generates 
a causet whose elements are the sprinkled points and whose 
order is that induced by the manifold's causal 
order restricted to the sprinkled points. We say that a causet $\CS$ 
is well approximated by a manifold $(\mathcal{M},g)$ if
it could have been generated, with relatively 
high probability, by sprinkling into $(\mathcal{M},g)$.

We propose the following definition of a discrete D'Alembertian,
$B$, on a causet $\CS$ that is a sprinkling, 
at density $\rho=l^{-4}$, into 4D Minkowski space
$\mbb{M}^4$.
Let $\phi:\CS\rightarrow\mbb{R}$ be a real scalar field, then
\begin{align}
B\phi(x)&:=\frac{4}{\sqrt{6}l^2}\big[-\phi(x)\nonumber \\
+&(\sum_{y \in L_1}-
9\sum_{y\in L_2}+16\sum_{y\in L_3}-8\sum_{y\in L_4})\phi(y)\big]\,,
\label{eqn2}
\end{align}
where the sums run over 4 \emph{layers} $L_i,\,i=1,\dots,4$, 
\be
L_i:=\{y\in \CS :\; y\prec x \; \text{and} \; n(x,y)=i-1\}
\ee
 and $n(x,y):= |I(x,y)|-2$. So, for 
example, layer $L_1$ is the set of all elements $y$ that are
linked to $x$ and as described above, they will be 
distributed close to a hyperboloid that asymptotes to the
past light cone of $x$ and is proper time $l$ away from $x$. 
This sum will not in general be uniformly convergent 
if it is over the elements 
of a sprinkling into infinite $\mbb{M}^4$ so we
introduce an IR cutoff, $L>>l$,
by embedding $\CS$ in $\mbb{M}^4$ and 
summing over the finitely many 
elements sprinkled in the intersection of the causal past of $x$ and
a ball of radius $L$ centred on $x$. 
The details of the calculation that shows why 4 layers are necessary
in 4D will appear elsewhere, however see \cite{Sorkin:2007qi}
for an explanation of why 3 layers are needed in 2D and
the conjecture that 4D will require 4 layers.

Now let $\phi$ be a real 
test field of compact support on $\mbb{M}^4$. 
If we fix a point $x\in \mbb{M}^4$ (which we always take to be included in $\CS$)
and evaluate $B\phi(x)$ on a sprinkling into 
$\mbb{M}^4$, its expectation value in this process 
is given by
\begin{align}
\bar{B}\phi&(x):=
\mbb{E}(B\phi(x)) = 
\fr{4}{\sqrt{6}l^2}\big[-\phi(x) \nonumber\\
&+\fr{1}{l^4}\int_{y\in J^-(x)}
                              \!\!\!\!\!\!\!\!\!\!\!d^4y\;\phi(y)\,
e^{-\xi}(1-9\xi+8\xi^2-\frac{4}{3}\xi^3)\big],
\label{eqn4}
\end{align}
where $\xi:=l^{-4}V(x,y)$, $V(x,y)$ is the volume of the 
causal interval between $x$ and $y$ and there is an implicit cutoff
$L$, the size of the support of $\phi$, on the integration range.

It can be shown that this mean converges, as the discreteness scale is
sent to zero, to the continuum D'Alembertian of $\phi$,
\be  
\lim_{l\rightarrow 0}\bar{B}\phi(x) = \Box\phi(x)
\ee
and that $\bar{B}\phi(x)$ is well approximated by 
$\Box\phi(x)$ when the characteristic length scale, $\lambda$,
on which $\phi(x)$ varies is large compared to $l$.
$\bar{B}$ is therefore effectively sampling the value of the
field only in a neighbourhood of $x$ of size
of order $l$ and the mean,
at least, of $B$ is about as local as
it can possibly be, given the discreteness.

To see roughly how this can happen, notice that 
the integrand in (\ref{eqn4}) is negligible for $\xi> \alpha^4$ where
 $\alpha$ is such that $e^{-\alpha^4} <<1$. The significant 
part of the integration range therefore
lies between the past light cone of $x$ and the hyperboloid 
$\xi = \alpha^4$ and comprises a part within 
a neighbourhood of $x$ of size 
$\alpha l$ -- whence the local 
contribution -- 
and the rest which stretches off far down the light cone. 
It is this second part of the range which
threatens to introduce nonlocality 
but because it can be coordinatized by $\xi$ itself and some 
coordinates $\eta^a$ on the hyperboloid the
integration over it will be proportional to
\be\label{suppress}
\int d^3\eta \int_0^{\alpha^4} d\xi e^{-\xi}(1-9\xi+8\xi^2-\frac{4}{3}\xi^3) \phi(\xi,\eta^a)
\,. 
\ee
If $\phi$ is nearly 
constant over length scale $\alpha l$, the $\xi$ integration is close to 
zero and the contribution is suppressed.
 
The fluctuations in $B\phi(x)$, 
however, are a different matter: if the physical 
IR cutoff $L$ is fixed and the discreteness scale sent to
zero, {\emph{i.e.}, the number of causet elements $N$
grows, simulations show the fluctuations around the mean grow
rather than die away and $B\phi(x)$ will not
be approximately equal to the continuum $\Box\phi(x)$.
To dampen the fluctuations we 
follow \cite{Sorkin:2007qi} and introduce an intermediate length scale 
$l_k\ge l$ and smear out the expressions 
above over this new scale, with 
the expectation that when $l_k >> l$ the inhering averaging will 
suppress the fluctuations via the law of large numbers. 
Thus we seek a discrete operator, $B_k$, whose mean is 
given by  (\ref{eqn4}) but with $l$ replaced by $l_k$:
\begin{align} 
\bar{B}_k\phi&(x)=\fr{4}{\sqrt{6}l_k^2}\big[-\phi(x)\nn
&+\frac{1}{l_k^4}\int_{y\in J^-(x)}
\!\!\!\!\!\!\!\!\!\!\!d^4y\;\phi(y)\,e^{-\xi}(1
-9\xi +8\xi^2-\frac{4}{3}\xi^3)\big], \label{meanbk}
\end{align}
where now $\xi:= l_k^{-4}V(x,y)$. 
Working back, one can show that
the discrete operator, $B_k$, with this mean is
\be
B_{k}\phi(x)=\frac{4}{\sqrt{6}l_k^2}\bigg[-\phi(x)+
\epsilon\sum_{y\prec x}f(n(x,y),\epsilon)\phi(y)\bigg],
\ee
where $\epsilon=(l/l_k)^4$ and
\begin{align}
f(n,\epsilon)=(1-\epsilon)^n 
&\left[ 
1-\frac{9\epsilon n}{1-\epsilon}
+\frac{8\epsilon^2n!}{(n-2)!(1-\epsilon)^2}\right.
\nn
-&\left.\frac{4\epsilon^3n!}{3(n-3)!(1-\epsilon)^3}
\right]
\,.
\end{align}
$B_k$ reduces to $B$ when $\epsilon =1$. 
$B_{k}$ effectively samples $\phi$ over elements in
 4 broad bands with a characteristic depth $l_k$, the bands' contributions
being weighted with the same set of alternating sign
coefficients as in $B$. 
Since (\ref{meanbk}) is just (\ref{eqn4}) with $l$ replaced by 
$l_k$, the mean of $B_{k}\phi(x)$ is close to $\Box\phi(x)$
when the characteristic scale over which $\phi$ varies is 
large compared to $l_k$. 
Now, however, numerical 
simulations show that the fluctuations 
are tamed. Points were sprinkled into a fixed causal interval
in $\mbb{M}^4$ between the origin and $t=1$ on the $t$ axis,
at varying density $\rho = \frac{N}{V}$,
where volume $V=\frac{\pi}{24}$. For each $N$,
100 sprinklings were done and for each sprinkling,
$B_k\phi$ was calculated at the topmost point 
of the interval for $\phi = 1$
and $l_k = 0.16$. For $N = 5000$, the mean was $\mu = 9,35$
and the standard deviation $s.d =134.8$. For $N = 10 000$,
$\mu = -4.00$ and $s.d.=102.6$ and for $N = 20 000$,
$\mu = 1.12$ and $s.d.=58.8$.
These results indicate that the fluctuations do die away, as
anticipated, as $N$ 
increases and are consistent with 
the dependence $N^{(-1/2)}$. Further results will appear elsewhere. 

The operators $B$ and $B_k$ derived in both 2D (in \cite{Sorkin:2007qi})
and 4D are defined in terms of the order relation on 
$\CS$ alone and so can be applied to a scalar field on 
{\textit{any}} causet. If, therefore, 
$(\mathcal{M},g)$ is a (2D or 4D) 
curved spacetime and $\phi$ is a scalar field 
on $\mathcal{M}$, we can compute $B_k\phi(x)$ on a sprinkling
into $\mathcal{M}$ and calculate its mean. 
Let $V_{2}$ and $V_{4}$ be the volumes of the intervals in 2D
and 4D respectively, $\xi_{2}:=V_{2}(x,y)l_k^{-2}$ and 
$\xi_{4}:=V_{4}(x,y)l_k^{-4}$. Then,
in the presence of curvature, 
\begin{align}
\bar{B}^{(2)}_{k}\phi(x)=\frac{2}{l_k^{2}}&\big[-\phi(x)+
\frac{2}{l_k^2}\int_{y\in J^-(x)}
\!\!\!\!\!\!\!\!\!\!\!d^2y\sqrt{-g}\;e^{-\xi_{2}}\nn 
&(1-2\xi_{2}+\frac{1}{2}\xi^2_{2})\phi(y)\big]
\end{align}
and
\begin{align}
\bar{B}^{(4)}_{k}\phi(x)=\fr{4}{\sqrt{6}l_k^2}
&\big[-\phi(x)+\frac{1}{l_k^4}\int_{y\in J^-(x)}
\!\!\!\!\!\!\!\!\!\!\!d^4y\sqrt{-g}\;\,e^{-\xi_{4}}\nn 
&(1-9\xi_{4}+8\xi_{4}^2-\frac{4}{3}\xi_{4}^3)\phi(y)\big],
\end{align}
in 2D and 4D respectively. 

These expressions can be evaluated using Riemann normal 
coordinates and in both cases we 
find 
\be
\lim_{l_k \rightarrow 0} \bar{B}^{(i)}_{k}\phi(x) 
= \left(\Box - \frac{1}{2} R(x)\right) \phi(x)\,.
\ee
The limit is a good approximation to 
the mean when the field $\phi$ varies slowly over 
length scales $l_k$ \emph{and} 
the radius of curvature $r>>l_k$. 

If the damping of fluctuations found in simulations in flat space are indicative
of what happens in curved space then, for a fixed 
large enough 
IR cutoff, $L$, the nonlocality 
length scale $l_k$ can be chosen such that $l << l_k << L$
and the value of 
$B_k\phi$ for a single sprinkling will be close to the mean. 
If $B_k$ is applied to the constant field $\phi= -2$, we therefore
obtain an expression that is close to the scalar curvature of the
approximating spacetime. 

In  each of 2D and 4D, we can now define a one parameter
family of candidate actions, $S_k[\CS]$, for a 
causal set, $\CS$, by summing $B_k(-1)$ over the elements
of $\CS$,
times $\hbar l^2$ to get the units right, times a number of 
order one which in 4D is the ratio of $l^2$ to $l_p^2$,  
where $l_p = \sqrt{8 \pi G \hbar}$ is the rationalized Planck length.  
When the nonlocality length 
$l_k$ equals the discreteness length $l$, $B_k = B$ 
and the action, $S[\CS]$ takes a particularly simple form 
as an alternating sum of numbers of small 
order intervals in $\CS$. 
Up to factors of order one, we have in 2D and 4D, respectively: 
\be \frac{1}{\hbar}S^{(2)}[\CS] = N - 2N_1 + 4N_2 - 2N_3
\ee
and 
\be \frac{1}{\hbar}S^{(4)}[\CS] = N - N_1 + 9N_2 - 16N_3 
+ 8 N_4\,,
\ee
where $N$ is the number of elements in $\CS$ and $N_i$ is 
the number of ($i+1$) element inclusive order intervals in  $\CS$.     

Because $B$ is the most non-nonlocal of the operators in the 
family, the action $S[\CS]$ is a sum of contributions each 
of which is not close to the value of the
 Ricci scalar at the corresponding
point of the continuum approximation. However, one might expect
that if the curvature is slowly varying on some intermediate scale, 
which we might as well call $l_k$, the averaging involved in the
summation  might perform the same role of suppressing 
the fluctuations as the smearing out of the 
operator itself so that the whole
action $S[\CS]$ is a good approximation to
the continuum action when $l_k$ is the appropriate size.

There are many new avenues to explore.
Can we use these results to 
define a quantum dynamics for causal sets? 
In 2D is there a relation with the Gauss-Bonnet 
theorem? Can we analytically continue the action in an 
appropriate way \cite{Sorkin:2009ka}
to enable Monte-Carlo simulations of the path sum?
What sort of phenomenology 
might emerge from such actions? 
To answer this latter question, we need to know how big 
 $l_k$ must be so that the action $S[\CS]$
is a good approximation to the Einstein-Hilbert action 
of the continuum $S_{EH}[g]$. In \cite{Sorkin:2007qi}, a rough
estimate is reported that in dimension 4, $l_k >> (l^2 L)^{1/3}$. 
Taking $L$ to be the Hubble scale, that
would mean that in the continuum regime, only
spacetimes whose curvature was constant over a scale $(l^2 L)^{1/3}$  
would be able to have an approximately local fundamental action. 
One might expect therefore that the 
phenomenological IR theory of
gravity that could emerge from such a fundamental theory would 
be governed by an effective Lagrangian
\be
\frac{{\cal{L}}_{\textrm{eff}}}
{\sqrt{-g}\hbar} = b_0 l_k^{-4} + b_1 l_k^{-2} R + b_2 R^2 + \dots
\ee
where $b_1$ and $b_2$ are of order 1, $b_0$ 
is set to its observed value, and where $l_k$ varies with epoch
and today is much larger than the Planck scale. The phenomenological 
implications of these ideas remain to be explored. 

We end by pointing out that these
results have a relevance beyond causal set theory 
as they provide a ``proof of concept'' 
for the mutual compatibility of Lorentz invariance, 
fundamental spacetime 
discreteness, and approximate locality.
\begin{acknowledgments}
We thank Rafael Sorkin for invaluable help and Michael 
Delph and  Joe
Henson for useful discussions. 
We also thank David Rideout for help with the simulations using his
CausalSets toolkit in the Cactus framework (www.cactuscode.org).
DMTB is supported by EPSRC. FD is supported by
by EC Grant No. MRTN-CT-2004-005616
and Royal Society Grant No. IJP 2006/R2.
We thank the Perimeter Institute for Theoretical Physics, Waterloo, 
Canada where much of this work was done. 
\end{acknowledgments}


\vspace{-0.5cm}

\bibliographystyle{apsrev}

\end{document}